\documentclass{nature}

\usepackage{amsmath}
\usepackage{amssymb}
\usepackage{hyperref}
\usepackage{capt-of}
\usepackage{multirow,booktabs}
\usepackage{url}
\usepackage{color,soul}  
\usepackage{ulem}
\usepackage{graphicx}
\usepackage{hanging}

\newcommand{\aap}{{Astron. Astrophys.}}

\newcommand{\aj}{{Astron. J.}}
\newcommand{\apj}{{Astrophys. J.}}
\newcommand{\apjl}{{Astrophys. J. Lett.}}
\newcommand{\apjs}{{Astrophys. J. Supp. Ser.}}

\newcommand{\araa}{{Annu. Rev. Astron. Astrophys.}}

\newcommand{\icarus}{{Icarus}}

\newcommand{\jgr}{{J. Geophys. Res.}}
\newcommand{\mnras}{{Mon. Not. Roy. Astron. Soc.}}
\newcommand{\na}{{New Astronomy}}
\newcommand{\nat}{{Nature}}

\newcommand{\okina}{\textquoteleft}
\newcommand{\OU}{{\okina}OUMUAMUA}
\newcommand{\Ou}{{\okina}Oumuamua}

\title{The Natural History of \Ou}

\author{The \Ou\ ISSI Team}

\begin{document}

\maketitle
\date{}

The discovery of the first interstellar object passing through the Solar System, 1I/2017 U1 (\Ou), provoked intense and continuing interest from the scientific community and the general public. The faintness of \Ou, together with the limited time window within which observations were possible, constrained the information available on its dynamics and physical state.  
Here we review our knowledge and find that in all cases the observations are consistent with a purely natural origin for \Ou. We discuss how the observed characteristics of \Ou\ are explained by our extensive knowledge of natural minor bodies in our Solar System and our current knowledge of the evolution of planetary systems. We highlight several areas requiring further investigation.

\section{What we Know about \Ou}
\label{sec:intro}

1I/\Ou\ was discovered on 2017 October 19 in the $w_{PS1}$-band observations of the PanSTARRS1 (PS1) Near Earth Object survey\cite{Meech:2017}.  \Ou\ was discovered three days after its closest approach to Earth at 0.16~au, well after it had passed closest to the Sun on 2017 September 9 at a perihelion distance of 0.25~au. By October 22 there was sufficient astrometry to securely identify that the orbit was hyperbolic\cite{Meech:2017}. Because of its rapid motion, there was only a short interval during which observations were possible. Within a week the brightness had dropped by a factor of 10 and within a month by a factor of 100.

The average brightness measured in visible wavelengths during the week after its discovery gave $H_V$=22.4\cite{Meech:2017,Jewitt:2017}, providing the first indication that \Ou\ has a radius in the hundred-meter range. {\sl Spitzer Space Telescope} observations in the infrared on November 21--22 did not detect \Ou\cite{Trilling:2018}. Their upper limits on the flux imply an effective radius between 49--220~m, depending on the assumed surface properties. For surface scattering parameters (called beaming parameters) that are typical of comets, this implies an effective radius of 70~m and a geometric albedo of 0.1. 
Relatively few minor bodies this small have been as well characterized physically, which hampers aspects of direct comparison of \Ou\ with similar objects from the Solar System. 

Several teams obtained photometric and spectral data in the optical to near-infrared to characterize \Ou's surface composition. 
\Ou\ is red, similar to many Solar System small bodies, e.g., comets, D-type asteroids, some Jupiter Trojans, and the more neutral trans-Neptunian objects\cite{Meech:2017,Jewitt:2017,Ye:2017,Bannister:2017,Fitzsimmons:2018,Bolin:2018}. 
Published measurements give a red slope at optical wavelengths of $\sim$10--20\%/100~nm. While the color is consistent with organic-rich surfaces, it is also consistent with iron-rich minerals, and with space weathered surfaces\cite{Moretti:2007}.  Thus, color alone is not diagnostic of composition.  Comparing the published spectroscopic and photometric data implies that some spectral variability with rotational phase is plausible within the data's uncertainties, but not certain\cite{Fitzsimmons:2018,Fraser:2018}. As albedo and spectral variability do not necessarily correlate, this does not imply any albedo variation, although it cannot be ruled out.

\Ou\ exhibited short-term brightness variation of over a factor of ten ($>$2.5 magnitudes)\cite{Meech:2017,Knight:2017,Jewitt:2017,Bannister:2017,Bolin:2018}.  The brightness range was unusually large. Of the minor planets in our Solar System with well-quantified light curves, there are only a handful of asteroids with brightness variations of this scale (\cite{Warner:2009}; last updated 31 January 2019). In most cases, these particularly high-amplitude light curves are based on observations of sub-$100$ m near-Earth asteroids at high phase angles, or on fragmentary light curves of slow-rotating objects. 

While brightness variations can be due to variations in the viewing geometry of a particular shape, or due to patchy albedo across a surface, minor planets' light curves are usually assumed to be shape-dominated, as their surfaces are thought to be covered by small regolith that is evenly distributed across the surface\cite{Fujiwara:2006}. \Ou's light curve shape, with narrow ``V-shaped'' minima and broad maxima, is indicative that its large brightness variations are caused by its shape, rather than variations in its albedo\cite{Lacerda:2007}.  Both phase angle and rotation state need to be considered in understanding \Ou's shape. Only a limited range of phase angles (19--27$^\circ$) could be observed in the short time span during which observations useful for defining \Ou's rotation were made. Accounting for the known effect of the enhancement of amplitude with increasing phase angle\cite{Zappala:1990}, the true ratio of longest axis to shortest axis was inferred to be $\geq$6:1\cite{McNeill:2018}. Due to the unknown orientation of \Ou's rotation pole, this axial ratio represents only a lower limit.

\begin{figure*}[t!]
\includegraphics[width=1.0\textwidth]{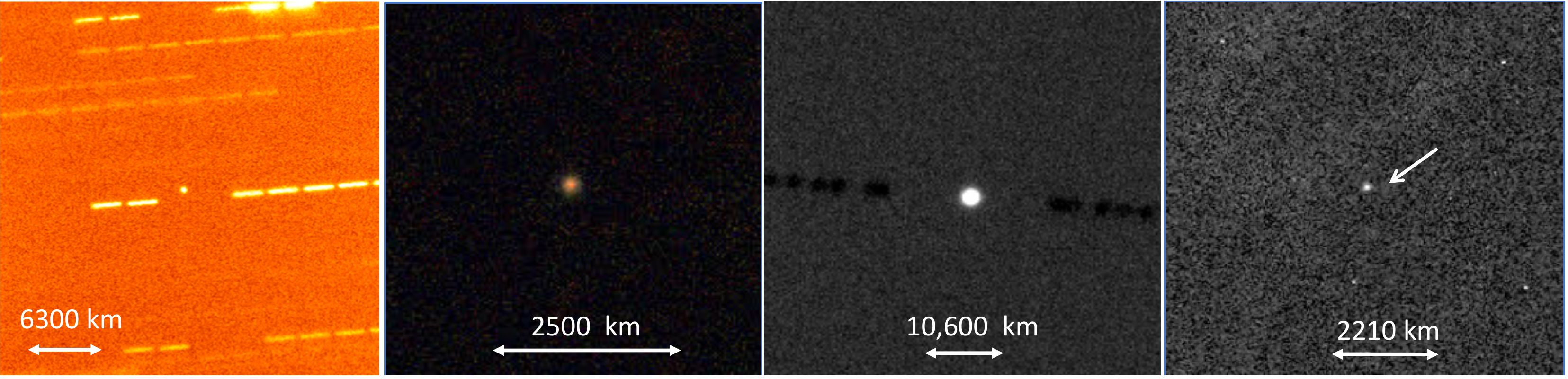}
\caption{
Montage of images of \Ou\ showing its point-like unresolved appearance with no hint of detectable activity.  From left to right: 0.4 hr integration through an R-band filter with the Nordic Optical Telescope on 2017 October 26\cite{Jewitt:2017}; ``true color'' image simulated from {\it grizY}-band images obtained on 2017 October 27 for a total integration of 1.6 hr with the Gemini South telescope\cite{Meech:2017}; a deep 3.6 hr $r$-band composite image obtained on 2017 October 27-28 with the Gemini North telescope\cite{Drahus:2018}; and an F350LP image from {\sl Hubble Space Telescope}\cite{Micheli:2018}.
}
\label{fig:res}
\end{figure*}

\Ou's brightness varied on a timescale of about 4 hours (implying a rotation period of $\sim$8 hours for a double-peaked lightcurve), but the various teams did not converge on a consistent rotation period while it was visible.  Analysis of the full photometric data set showed that \Ou\ was in a state of excited rotation\cite{Fraser:2018,Drahus:2018,Belton:2018}.  The most comprehensive model published to date\cite{Belton:2018} concluded that \Ou\ is rotating around its shortest axis with a period of 8.67$\pm$0.34 hours, and has a likely period of rotation around the long axis of 54.48 hours. How we interpret the shape of \Ou\ depends on its specific state of rotation, including its rotation pole.  \Ou\ can either have a narrow elongated-ellipsoid shape or a shape more reminiscent of a flattened oval.

Sensitive searches for activity (Fig.~\ref{fig:res}) showed no evidence for micron-sized dust near \Ou\cite{Meech:2017,Jewitt:2017,Ye:2017,Micheli:2018}. However, the observations were not sensitive to the detection of millimeter-sized and larger dust, so we have no constraints for the presence of large grains. There was also no detection of any gas, including searches for CN, H$_2$O, CO and CO$_2$\cite{Ye:2017,Fitzsimmons:2018,Park:2018,Trilling:2018}, although the level to which each gas can be ruled out varies significantly. We summarize these and other measured properties of \Ou\ in Table~\ref{tbl:property}.

A detailed investigation of the astrometric position measurements from the first observations in mid-October 2017 through the last observations obtained by the {\sl Hubble Space Telescope} on January 2, 2018, showed that a gravity-only orbit provided an inadequate fit to the data. Instead, the data were well fit with the addition of a radial acceleration varying as 1/$r^2$, where $r$ is the heliocentric distance\cite{Micheli:2018}. This type of acceleration is frequently used in orbital studies of comets, and usually interpreted as being due to an activity-driven cometary acceleration consistent with the decreasing energy with distance from the Sun. 

\begin{table*}
\begin{center}
\caption{A summary of measured properties of \Ou\ \label{tbl:property}}
\begin{tabular}{lllc}
\hline
Quantity & & Value & References\\
\hline
\multicolumn{3}{l}{\bf Dynamical Properties} \\
Perihelion distance    & $q$ [au]                         & $0.255912\pm0.000007$                     & [1] \\
Eccentricity           & $e$                              & $1.20113\pm0.00002$                       & [1] \\
Incoming radiant       & $\alpha$, $\delta$ [deg]         & 279.4752, 33.8595                         & [2] \\
Earth close approach   & $\Delta$ [au]                    & $0.16175\pm0.00001$                       & [1] \\
Incoming velocity      & $v_{\infty}$ [km s$^{-1}$]       & $26.4204\pm0.0019$                                   & [2] \\
Non-grav acceleration  & $A_1 r^{-2}$ [m s$^{-2}$] & (4.92$\pm$0.16)$\times 10^{-6}$     & \cite{Micheli:2018} \\       
\hline
\multicolumn{3}{l}{\bf Physical Properties} \\
Absolute magnitude    & $H_V$                             & $22.4\pm0.04$                             & \cite{Meech:2017} \\
Albedo                & $p_V$                             & $>$ [0.2,0.1,0.01]                        & \cite{Trilling:2018} \\
Effective diameter    & $D_N$ [m]                         & $<$[98,140,440]                & \cite{Trilling:2018} \\
Rotation state        &                                   & complex, long-axis mode                              & \cite{Fraser:2018,Belton:2018,Drahus:2018} \\ 
Rotation period       & $P$ [hr]                          & $8.67\pm0.34$~hr (long-axis precess)      & \cite{Belton:2018} \\
Axis ratio            & a:b                               & $>$6:1                                    & \cite{McNeill:2018} \\
Shape                 &                                   & cigar, or oblate spheroid                 & \cite{Belton:2018} \\
Spectral slope        & $S_V$ [\% per 100 nm]             & 23$\pm$3, 10$\pm$6, 9.3--17              & \cite{Meech:2017,Ye:2017,Fitzsimmons:2018} \\ 
Surface spectral type &                                   & D-type                                & \cite{Meech:2017,Fitzsimmons:2018} \\
H$_2$O production     & $Q$(H$_2$O) [molec s$^{-1}$]      & 4.9     $\times$10$^{25}$ @ 1.4 au (model)& \cite{Micheli:2018} \\
OH production         & $Q$(OH) [molec s$^{-1}$]          & $<$ 1.7 $\times$10$^{27}$ @ 1.8 au (obs)  & \cite{Park:2018} \\
Hyper volatile (CO?)  & $Q$(X) [molec s$^{-1}$]           & 4.5     $\times$10$^{25}$ @ 1.4 au (model)& \cite{Micheli:2018} \\
CO$_{2}$ production   & $Q$(CO$_2$) [molec s$^{-1}$]      & $<$ 9   $\times$10$^{22}$ @ 2.0 au (obs)  & \cite{Trilling:2018} \\
CO production         & $Q$(CO) [molec s$^{-1}$]          & $<$ 9   $\times$10$^{23}$ @ 2.0 au (obs)  & [3]\\
CN production         & $Q$(CN) [molec s$^{-1}$]          & $<$ 2   $\times$10$^{22}$ @ 1.4 au (obs)  & \cite{Ye:2017} \\
C$_2$ production        & $Q$(C$_2$) [molec s$^{-1}$]          & $<$ 4   $\times$10$^{22}$ @ 1.4 au (obs) & \cite{Ye:2017} \\
C$_3$ production        & $Q$(C$_3$) [molec s$^{-1}$]          & $<$ 2   $\times$10$^{21}$ @ 1.4 au (obs)  & \cite{Ye:2017} \\
Dust production       & $Q$(dust) [kg s$^{-1}$]           & $<$ 1.7 $\times$10$^{-3}$ @ 1.4 au (obs)  & \cite{Meech:2017} \\
                      &                                   &
$<$ 10 @ $\sim10^3$ au (obs)  & [3] \\
\hline
\end{tabular}
\end{center}
$^{\dag}$Reference Key:  [1] JPL Horizons orbital solution \#16; [2] \cite{Bailer-Jones:2018} using the pure $1/r^{2}$ radial acceleration solution from \cite{Micheli:2018}; [3] M.\ Mommert (priv.\ comm.) revising the calculation in \cite{Trilling:2018}.
\end{table*}

\section{A Critical Review of Current Theories}
\label{sec:theories}

The detection of interstellar objects was anticipated for decades\cite{McGlynn:1989} due to our understanding of how planetary systems form and evolve, but \Ou\ managed to surprise us nonetheless.  Most notably, it was generally assumed that the first interstellar object would be an obviously active comet because they are much brighter than an asteroidal object for a given nucleus size.  The assumption seemed natural because of the  expected similarity between interstellar interlopers and objects from the Solar System's Oort cloud that have been stored in the deep-freeze of deep-space for billions of years.  Since it was thought that most objects from the Oort cloud appear as long period comets, the interstellar objects were expected to have the same morphology.  Thus, with limited exceptions\cite{Engelhardt:2017}, most speculation on the properties and discovery of interstellar objects involved strongly active comets.  The belief that most Oort cloud objects become active comets when they enter the inner solar system drove most of the limits on the spatial density of interstellar objects.  We now know that there are many inactive or weakly active Oort cloud objects\cite{Meech:2016} and if we assume that interstellar objects share the same characteristics then their spatial density would be higher than originally expected.  The second surprising aspect of \Ou\ is that it was discovered much faster than expected --- early predictions were that PanSTARRS1 was unlikely to detect an interstellar object in 10 years of operation\cite{Jewitt:2003}.  Finally, \Ou\ is much smaller than would have been imagined as the expectation was that it would be similar to a long period comet, a km-scale active object.  Detecting and characterizing objects within our own Solar System of \Ou's size is limited to near-Earth objects, so that even if its size had been anticipated there are limited examples from which to form our expectations.  Furthermore, even if its small size had been predicted, with a corresponding likelihood of being irregularly shaped, no one would have imagined it to be so unusually elongated.  Despite all these surprises \Ou's properties can be readily and naturally explained.

\subsection{\Ou\ Originated in a Planetary System}
\label{sec:origin}

\begin{figure*}[t!]
\centering
\includegraphics[width=0.9\textwidth]{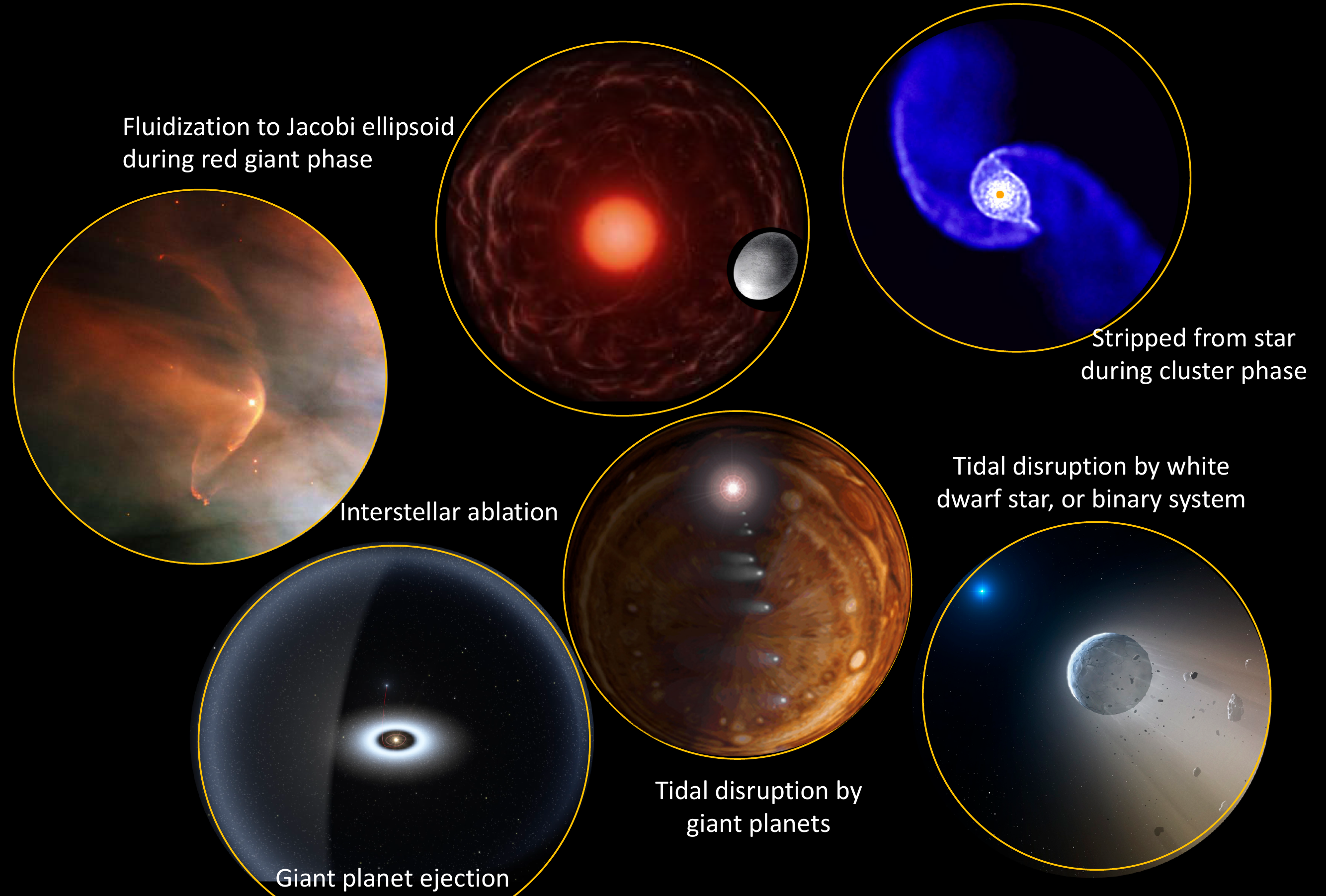}
\caption{
Montage of potential formation scenarios of \Ou\ as a natural planetesimal. 
}
\label{fig:form}
\end{figure*}

A number of processes have been invoked to explain \Ou's origins and peculiarities since its discovery (Fig.~\ref{fig:form}).  These models generally expect \Ou\ or its parent body to have been born as a planetary building block -- a planetesimal -- in a gas-dominated protoplanetary disk around a young star. Planetary disks containing planetesimals are common around very young stars ($<$3 Myr\cite{Haisch:2001,Pfalzner:2014}). Roughly 20\% of slightly older Sun-like stars are observed to still have mid-infrared excess emission\cite{Montesinos:2016}, interpreted as the dust generated by colliding outer planetesimals (``debris disks''\cite{Wyatt:2008}). This implies that a large fraction of stars are indeed born with large reservoirs of planetesimals capable of being dynamically ejected.

A straightforward explanation for \Ou\ is that it is a planetesimal (or a planetesimal fragment) ejected from its home system\cite{Raymond:2018,Raymond:2018b}. During planetary system formation, a significant portion of a system's planetesimals are ejected into interstellar space\cite{Charnoz:2003}.
Gravitational interactions with the stars of the surrounding cluster or with the giant planets of the planetary system itself are major mechanisms of ejection\cite{Tremaine:1993}.
Simulations show that planetesimals are most efficiently ejected in systems in which the giant planets themselves become unstable\cite{Raymond:2010}.  
In close binary systems (with a planet-forming disk exterior to two stars), planetesimals that enter within a critical distance to the binary are destabilized\cite{Holman:1999} and quickly ejected as interstellar objects\cite{Jackson:2018}.  Close stellar flybys, which are common during the $\sim 3-5$ Myr-long embedded cluster phase\cite{Vincke:16},
can strip planetesimals from the outer parts of planetary systems\cite{Hands:2019}.
As their host stars evolve off the main sequence and lose mass, planetesimals will eventually also be liberated from their home systems\cite{Veras:2016}.

\subsection{The Expected Number Density of Interstellar Objects in Space}
\label{sec:numdensity}

Combining the observed absolute magnitude of \Ou\ with current sky-survey detection limits, the number density of objects in interstellar space of the same size as \Ou\ or larger is about 0.1 per cubic au\cite{Meech:2017,Trilling:2017,Do:2018}. This estimate applies to objects with little to no activity (like \Ou) and implies that interstellar objects are continuously passing through the Solar System below our current detection threshold.  

It has been asserted that this number density of interstellar objects is 2--8 orders of magnitude higher than would be expected from planet formation scenarios\cite{Bialy:2018}. However, transforming a number density of interstellar objects to a mass density requires a knowledge of the population's size-frequency distribution (SFD)\cite{MoroMartin:2009}. With a single detected object there are no firm constraints on this distribution: until the interstellar object SFD is known from tens of detections, there is a disconnect between the measured number density of interstellar objects and their mass density. 

\begin{figure*}[t!]
\centering
\includegraphics[width=0.75\textwidth]{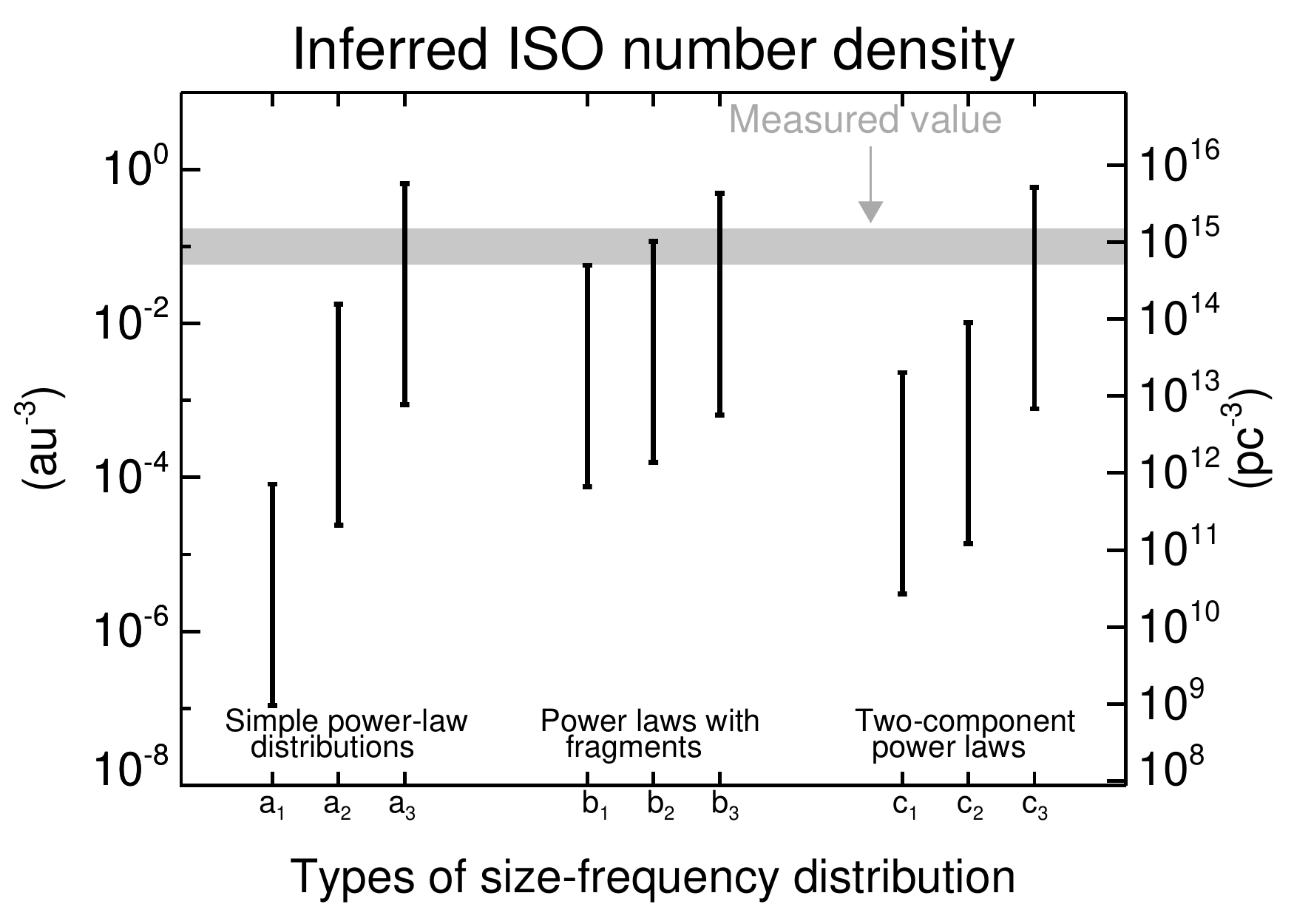}
\caption{Inferred number density of interstellar objects -- for a fixed estimate of the mass density of 0.004--3 Earth masses per cubic parsec -- assuming different underlying size-frequency distributions (SFDs).  We tested three SFDs: 1) power laws ($a_{1-3}$) in which the number of objects $N$ of a given mass $m$ is $N(m)\propto m^{-x}$; 2) power laws in which a small fraction (typically 1\%) of the mass has been converted into fragments --  comparable in size to \Ou, perhaps due to tidal disruption from giant planet encounters prior to ejection\cite{Raymond:2018,Raymond:2018b} ($b_{1-3}$); and 3) two-component power laws ($c_{1-3}$). The power laws extend from effective radii $r_{min}$ to $r_{max}$ with $N(m)\propto m^{-x}$, and all three have $r_{max}=100$~km. Distribution $a_1$ is consistent with simulations of planetesimal formation\cite{Simon:2017,Schafer:2017} and has $r_{min}=100$~m and $x=0.6$. Distribution $a_2$ assumes collisional equilibrium\cite{Dohnanyi:1969} and has $r_{min}=50$~m and $x=5/6$. Distribution $a_3$ is bottom-heavy (the smallest objects dominate by mass); it extrapolates the size-frequency distribution of boulders on comet 67P/Churyumov-Gerasimenko\cite{Pajola:2015} to large sizes and has $r_{min}=50$~m and $x=1.2$. Distribution $b_1$ contains 99\% of its mass following distribution $a_2$ with 1\% by mass in 50 m-sized fragments. Distribution $b_2$ contains 97\% of its mass following distribution $a_1$ and 3\% in 50 m-sized fragments (see \cite{Raymond:2018}). Distribution $b_3$ is a single-size distribution, assuming that all interstellar objects are \Ou-sized (100~m). Distributions $c_1$ through $c_3$ all assume $r_{min}=50$~m and $r_{max}=100$~km. Distribution $c_1$ has $x=0.6$ for objects larger than $r_{break}=1$~km and $x=5/6$ for smaller ones. Distribution $c_2$ has the same power laws but with $r_{break}=10$~km. Distribution $c_3$ has $x=0.6$ for objects larger than $r_{break}=10$~km and $x=1.2$ for smaller ones.
}
\label{fig:num_density}
\end{figure*}

We show with a simple experiment that the expected number density of interstellar objects varies by many orders of magnitude depending on the SFD applied to the mass (Fig.~\ref{fig:num_density}).
Our estimate is based on the idea that \Ou\ is a planetesimal (or a planetesimal fragment) that was ejected from its home system by giant planets\cite{Raymond:2010,Raymond:2018}.

We first estimate the underlying mass density of interstellar objects based on planet formation theory and observational constraints.
The density and mass distribution of stars are  well-known\cite{Kroupa:1993};
they are dominated by low-mass stars, with a Galactic disk-averaged value of $\sim$0.2 stars per cubic parsec. Virtually all stars host planets\cite{Cassan:2012}. Radial velocity surveys find that $\sim$10--20\% of Sun-like stars have gas giants\cite{Mayor:2011} but this fraction drops significantly for low-mass stars\cite{Johnson:2007}.
The stellar mass-averaged frequency of gas giants is $\sim$1--10\%\cite{Winn:2015}. Microlensing surveys find that the occurrence rate of ice giants is significantly higher ($\sim$10--50\%) and has a weaker stellar mass dependence\cite{Suzuki:2016}.
Similarly, the ubiquity of gap structures in the ALMA disks suggests that Neptune-mass planets are common at large distances, with an occurrence rate estimated at $\sim$50\%\cite{ZangS:2018}.

How much mass in planetesimals does each system eject?  
This depends on the dynamics of each individual system and whether the planets remain stable\cite{Raymond:2010}. 
We assume that each gas giant system ejects 1--100 Earth masses\cite{Raymond:2010}. The abundant ice giants also efficiently eject planetesimals during\cite{Izidoro:2015} and after\cite{Tremaine:1993} their formation; we assume each ice giant system ejects 0.1--10 Earth masses.
Allowing for the frequency of the types of planetary systems, this comprises 0.1--10 Earth masses per star ejected by gas giants and 0.01--5 Earth masses by ice giants. 
This totals 0.02 to 15 Earth masses in interstellar objects per star or 0.004 to 3 Earth masses per cubic parsec.

We then calculate the expected number density of interstellar objects from that mass density estimate.
Figure~\ref{fig:num_density} shows the huge diversity of number densities of interstellar objects that can be inferred: the differences arise purely from the choice of plausible size-frequency distribution.  While the uncertainty in our estimate of ejected planetesimal mass per star spans three orders of magnitude, the difference in inferred number density between SFDs is even larger.  For example, a power-law distribution characteristic of planetesimal formation simulations (SFD $a_1$) requires an implausibly large amount of mass -- thousands of Earth masses -- to be ejected per star in order to match the observational constraint on the number density\cite{Raymond:2018,Rafikov:2018b,MoroMartin:2018,MoroMartin:2019a}. However, several SFDs from Fig.~\ref{fig:num_density} with somewhat more mass in small objects (e.g., SFD $b_2$ has 3\% by mass in fragments and is otherwise similar to SFD $a_1$) can match the measured interstellar object number density. It is easier to match the inferred number density at the higher end of our estimate of the interstellar object mass density, but the main uncertainty comes from the assumed SFD.

Thus, given that the number density of interstellar objects cannot yet be reliably related to the mass density, the claim that the observed number density is presently ``higher than expected'' from planet formation scenarios is not supported.

\subsection{Uniqueness of the Trajectory} 
\label{sec:trajectory}

While not typical for field stars, \Ou's trajectory is exactly what was expected for detectable interstellar objects\cite{Engelhardt:2017}.  As they age, stars in the solar neighbourhood are perturbed away from the Local Standard of Rest, which is defined by the galactic motions of nearby stars.
Of course, a small fraction of older stars may still have small random velocities\cite{Burgasser:2015}. \Ou's random velocity is 9~km~s$^{-1}$ from the Local Standard of Rest, far smaller than the $\sim$50~km~s$^{-1}$ velocity dispersion of nearby stars\cite{Anguiano:2017}. This small random velocity could imply that \Ou\ is dynamically young\cite{Meech:2017}, with a statistically-derived dynamical age of $<$2 Gyr\cite{Portegies:2018,AlmeidaFernandes:2018}.

\begin{figure*}[t!]
\centering
\includegraphics[width=1.0\textwidth]{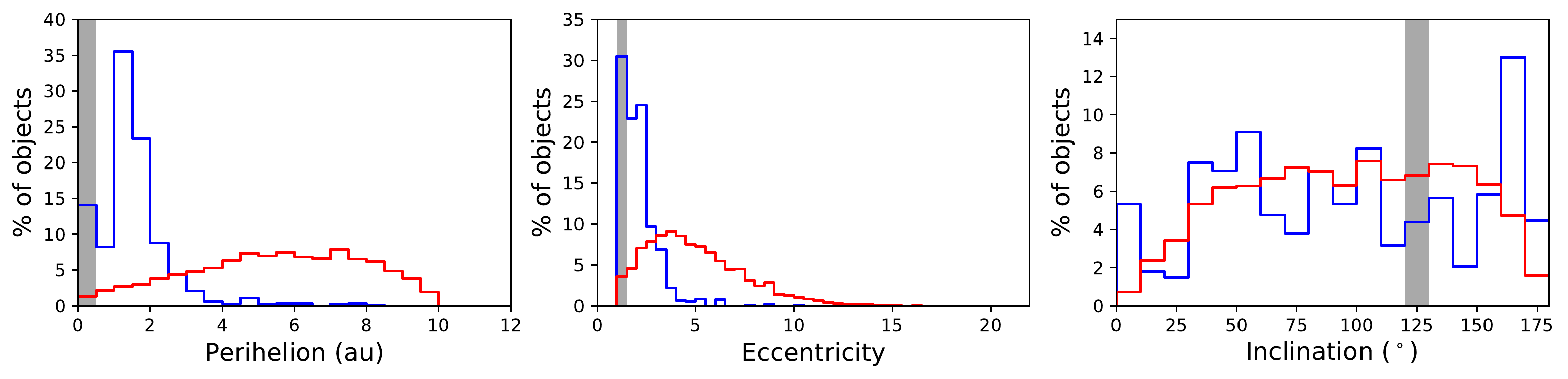}
\caption{
Predicted distribution of orbital elements of natural interstellar objects (blue curves are inactive objects, red curves are active objects) detected by the primary contemporary asteroid surveys (adapted from \cite{Engelhardt:2017}).  In each distribution, \Ou\ (gray vertical bar) has orbital elements at or near the most likely orbital elements for inactive objects. 
}
\label{fig:elements}
\end{figure*}

Gravitational focusing by the Sun creates an observational bias that favors the detection of interstellar objects with low random velocities, like that of \Ou\cite{Engelhardt:2017}.  This means it is challenging to use \Ou's galactic motion to constrain the interstellar object population's velocity dispersion.  Indeed, as demonstrated in Figure~\ref{fig:elements}, there appears to be nothing unusual about the specific parameters of \Ou's hyperbolic trajectory, as its perihelion distance, eccentricity, and inclination agree well with the predicted distribution of the values for interstellar objects detectable by the major contemporary asteroid surveys --- a prediction published\cite{Engelhardt:2017} nearly eight months {\it before} \Ou\ was discovered!

\subsection{``Cometary'' Activity and Retention of Volatile Materials}

The mass loss needed to explain \Ou's observed non-gravitational acceleration\cite{Micheli:2018} is on the order of 1 kg~s$^{-1}$. Outgassing models for an object the size of \Ou\ with comet-like properties can produce this amount of mass loss at the distances observed\cite{Meech:2004}. 
Furthermore, when the {\it Rosetta} observations of comet 67P/Churyumov-Gerasimenko (made at comparable heliocentric distances to when \Ou\ was observed) are scaled down to an \Ou-sized object, they yield a similar outgassing rate\cite{Paetzold:2018}. Depending on the assumptions, the total mass lost during the interval of observations may represent up to $\sim$10\% of {\Ou}'s total mass\cite{Seligman:2019}.

A typical comet with this level of outgassing would produce dust of all sizes, yet no dust was detected. The absence of a radiation-pressure-swept tail indicates that if any particles were released, the effective particle size must be large. Observations by both ground-based telescopes and space missions to comets have shown that the ejection of fine-grained dust, which dominates the reflected light at visible wavelengths, is not always correlated with gas release. For example, comet 2P/Encke reaches a similar perihelion distance as \Ou, and it often lacks any detectable dust at visible wavelengths\cite{Fernandez:2000}.
Some long-period comets preferentially eject large particles due to a mechanism that is currently not understood\cite{Sekanina:1982}.
Unfortunately, no observations were sensitive to large dust grains, which are most detectable in radio wavelengths, and meteor observations (sensitive to 0.1--1~mm dust) can only rule out activity at unrealistically large heliocentric distances ($>1000$~au) or with unusual strength\cite{Ye:2017}.

The search for gas emission from \Ou\ was not comprehensive owing to the challenging observing circumstances.  There were no observations that could have made sensitive enough detections of water outgassing to test for comet-like activity.  The relative abundance of CN to H$_2$O of \Ou\ needed to reconcile the non-detections of CN\cite{Ye:2017} with the inferred H$_2$O outgassing rate needed to account for non-gravitational forces\cite{Micheli:2018}, while unusual, is not unprecedented. \Ou\ needed to be depleted in CN by at least a factor of 15 relative to typical abundances in comets, while comets C/1988 Y1 Yanaka\cite{Fink:1992} and 96P/Machholz~1\cite{Schleicher:2008} 
were depleted by factors of 25 and 72, respectively. One of these highly depleted comets, 96P, also has a very low amount of dust observed at visible wavelengths compared to gas\cite{Schleicher:2008}, like \Ou.

The upper limits to the CO and CO$_2$ production rates\cite{Trilling:2018}  (and M.\ Mommert priv.\ comm.) combined with the inferred H$_2$O production rate imply abundances of CO/H$_2$O $\leq$ 2\% and CO$_2$/H$_2$O $\leq$ 0.2\%. This CO upper limit is within the range of measurements for known comets, while the CO$_2$ upper limit is about an order of magnitude lower\cite{Ahearn:2012}. However, CO$_2$ is difficult to measure, so the known sample may be biased to higher abundances. Both CO and CO$_2$ are much more volatile than H$_2$O, resulting in a trend to lower ratios with smaller heliocentric distances\cite{Ootsubo:2012}. The volatility difference would have resulted in CO and CO$_2$ being depleted deeper than H$_2$O. Thus, \Ou\ may have lost most/all of its CO and CO$_2$ prior to the observations that would have constrained their abundances. Alternatively, it may have had intrinsically low abundances of CO and CO$_2$ due to formation conditions in its home system. The range of these ratios of volatiles in comets has recently been found to be far greater than was previously known: C/2016 R2 PanSTARRS has CO/H$_2$O at least several orders of magnitude higher than any other measured comet, with no H$_2$O yet conclusively detected\cite{Biver:2018}.

Thermal models show that ices may exist within just $\sim 30$~cm of the surface without being released during \Ou's perihelion passage\cite{Fitzsimmons:2018,Seligman:2018}.  A natural consequence would be a thermal lag in which outgassing begins significantly later. Such a scenario would decrease the total amount of volatile material needed to explain the observed non-gravitational acceleration and shorten the timescale over which torques were at work. One thermal model was shown to be consistent with the observed non-gravitational acceleration by assuming outgassing from water in combination with another volatile species\cite{Micheli:2018}.

Based on the lack of detected activity, it has been suggested that \Ou\ had repeated passages close to its host star before being ejected\cite{Raymond:2018b}. Such repeated close passages can remove volatiles from planetesimals' surfaces and render ejected planetesimals inactive, or extinct\cite{Rickman:1991}. Models that match the various distributions of Solar System comets\cite{Nesvorny:2017} predict that smaller objects become inactive more quickly, so it could simply be that 100-m scale ejected objects like \Ou\ are devolatilized in their outer layers.  There could very well be a population of inactive small objects from our own Oort cloud that goes undetected because of their lack of activity, as evidenced by the Manx objects\cite{Meech:2016}.

Besides outgassing, a number of possible explanations for the observed non-gravitational acceleration were considered, but ultimately rejected\cite{Micheli:2018}. Most prominently was solar radiation pressure, which required {\Ou}'s density to be 3--4 orders of magnitude lower than that of asteroids of similar size (solar radiation pressure effects have been detected on a few small asteroids\cite{Micheli:2012}). 
Alternate explanations in support of solar radiation pressure have suggested that \Ou\ had a low density due to a fractal aggregate structure produced either by devolatilization of a comet-like body prior to its discovery\cite{sekanin:2018} or having formed as a very large aggregate of icy dust particles beyond the snow line in its home system\cite{MoroMartin:2019b}. Such extended, extremely low density objects have never been detected, but might naturally explain some other phenomena observed for disrupting comets\cite{sekanin:2018} or help reconcile some aspects of protoplanetary disc models\cite{MoroMartin:2019b}.

\subsection{Alien technology?}
\label{sec:solarsail}
The idea of \Ou\ as alien technology has been advocated in a series of papers\cite{Loeb:2018,Bialy:2018,Siraj:2019}. The authors argue that the dimensions needed to explain the observed solar radiation pressure are consistent with a ``solar sail.'' While this fits some aspects of the observations --- the basic idea of \Ou\ having a highly flattened shape was previously considered\cite{Belton:2018,Micheli:2018} --- it appears unable to explain other key aspects of the observations, and some arguments in favor of this hypothesis are simply wrong. 

The key argument against the solar sail hypothesis is \Ou's light curve amplitude. In order for a solar sail to cause the observed non-gravitational acceleration, it needs to remain properly oriented towards the Sun. However, in order to yield the observed brightness variations, its orientation would need to be varying as viewed from Earth. Furthermore, since the actual dimensions of the solar sail would be $>10:1$, the orientation as viewed from Earth would need to be very nearly edge on, and remain so throughout the observations despite viewing geometry changes. It has not been shown that an orientation exists that can achieve all of these constraints imposed by the observational data.
Furthermore, as discussed earlier, the shape of \Ou's light curve, with broad maxima and narrow minima, is consistent with an elongated ellipsoid.  

The claim\cite{Loeb:2018} that \Ou\ must be at least ten times ``shinier'' than all Solar System asteroids to make the {\sl Spitzer Space Telescope} data consistent with the ground based observations is incorrect. The {\sl Spitzer} observations are consistent with geometric albedos $0.01\leq p_v\leq 0.5$\cite{Trilling:2018}, with a most likely albedo of $p_v \sim 0.1$. Comets have geometric albedos of $p_v=0.02-0.07$, carbonaceous and silicate asteroids have $p_v=0.05-0.21$, and the most reflective asteroids have $p_v\sim 0.5$\cite{Thomas:2011,Kokotanekova:2017}.
Thus \Ou's measured reflectivity of $\sim0.1$ is entirely consistent with normal Solar System small bodies.

Finally, it was argued that \Ou\ was deliberately sent toward Earth based on its ``unusual'' kinematics and presumed scarcity\cite{Bialy:2018}. While provocative, this argument is baseless. First, \Ou's trajectory is consistent with predictions\cite{Engelhardt:2017} for detectable inactive interstellar objects. Second, the measured number density cannot be claimed to be at odds with expectations because of our ignorance of the size distribution of interstellar objects.   

Thus, we find no compelling evidence to favor an alien explanation for \Ou.

\section{Open questions}
\label{sec:questions}

We have discussed the many aspects of \Ou's properties that can be explained naturally.
However, there remain several unanswered questions regarding \Ou\ that warrant further study.

\subsection{Shape}
\label{sec:shape}

While several models have been proposed to explain \Ou's very elongated shape, none can naturally match such an extreme axis ratio (of at least 6:1) within a self-consistent framework. 
One model\cite{Katz:2018} invokes the complete fluidization of a planetesimal by an evolving red giant star, causing the object to assume the shape of a high angular momentum Jacobi ellipsoid. Other models have proposed that \Ou\ is a fragment of a planetesimal\cite{Raymond:2018b,Rafikov:2018,sekanin:2018} or planet\cite{Cuk:2018} that was tidally disrupted after a very close passage to a low-mass star, white dwarf, or giant planet, or simply as it neared perihelion. It remains to be demonstrated whether such disruption events create fragments as stretched-out as \Ou\ appears to be.  A third model proposes that a large number of high-velocity impacts with dust grains may create sharp edges and planar surfaces on small bodies\cite{Domokos:2009} or simply erode enough material to substantially increase the axis ratio of small objects\cite{Vavilov:2019}, while a fourth proposes that it formed from a low speed collision between two $\sim$50 m planetesimals in a protoplanetary disk\cite{Sugiura:2019}. In the context of these models, it remains to be understood why such extreme shapes are so rare among larger Solar System bodies; though this may partly be an observational selection effect. At two orders of magnitude larger than \Ou, the primordial Kuiper belt object 2014 MU$_{69}$ has a bi-lobed structure with substantive ``pancake" flattening to the larger lobe\cite{Stern:2019}.

\subsection{Rotation state}
\label{sec:rotation}

The ensemble of published photometry reveals that \Ou\ is in non-principal axis rotation (NPA)\cite{Fraser:2018,Drahus:2018,Belton:2018}, which is a spin state commonly observed among asteroids, including objects as small as \Ou\cite{Pravec:2005}. The details of the NPA are non-unique from the available data, including when \Ou\ achieved NPA rotation. Disruption or strong gravitational encounters could have created the NPA state, and the $>10^{11}$~yr damping timescale is sufficiently long that the tumbling may have originated in or during departure from its home system\cite{Fraser:2018,Drahus:2018,Belton:2018,Kwiecinski:2018}. Alternatively, the NPA rotation might have occurred during \Ou's journey through our system. It has been argued that the level of outgassing needed to explain the non-gravitational acceleration would have resulted in a rapid change in rotation period\cite{Rafikov:2018}. Even a small asymmetry in the torquing might have perturbed \Ou\ from simple rotation to NPA rotation.

One work found that if the large non-gravitational acceleration was caused by typical cometary outgassing, then the associated torques should have caused \Ou\ to rapidly spin up beyond its rotational break up limit\cite{Rafikov:2018}. In contrast, others showed that outgassing activity that followed the subsolar point of an elongated body could produce the observed non-gravitational acceleration and would naturally result in NPA rotation with a lightcurve amplitude and period comparable to the observations, without causing extreme spin up\cite{Seligman:2019}.

The orientation of \Ou's rotational angular momentum vector is unconstrained from the finite available data, but is critical for properly assessing the shape from the light curve. Dynamical work found that the rotation can be in one of five different modes, and if it is closest to its lowest rotational energy the shape can resemble the elongated ``cigar-like'' shape, and only in the highest energy state would it be an ``extremely oblate spheroid''\cite{Belton:2018}.   The ``cigar-like'' shape is the more likely configuration, both because it is energetically more stable and because it permits a much larger range of orientations on the sky (as previously discussed, a very flat shape requires a very specific orientation to produce the observed light curve).

\subsection{Home system}
In spite of many attempts to trace the orbit of \Ou\ back to its home system\cite{Bailer-Jones:2018,Zuluaga:2018,Dybczynski:2018,Zhang:2018} or star cluster\cite{Gaidos:2018,Feng:2018}, no convincing candidate origin star systems or stellar associations have been identified.  Whether tracing back to a unique origin is feasible depends on how long ago \Ou\ was ejected from its home system, since more distant regions must be considered for longer travel times, and whether it had past encounters, since each effectively erases its dynamical past. Although future data releases of high precision surveys like Gaia are likely to spur deeper searches and may yet reveal plausible candidates, it is likely that no system will be definitively shown to be \Ou's origin. 

In addition to travel time, uncertainties in velocity/acceleration affect our ability to identify its home system. The first generation of searches\cite{Gaidos:2018,Feng:2018,Zuluaga:2018,Dybczynski:2018,Zhang:2018} were based on the Keplerian orbit solution available at the time, while a later study\cite{Bailer-Jones:2018} utilized the solution that included non-gravitational acceleration, assuming that it was symmetric pre- and post-perihelion. Whether this assumption is justified is ultimately unknown as no pre-perihelion observations are available, but it is likely that outgassing was delayed due to a thermal lag\cite{Fitzsimmons:2018}.
Without observational constraints, the parameter space to search for a home system increases considerably.

\section{Conclusions}
As the first interstellar visitor to our solar system, \Ou\ has challenged many of our assumptions about how small bodies from another star system would look. While \Ou\ presents a number of compelling questions, we have shown that each can be answered by assuming \Ou\ to be a natural object. 
Assertions that \Ou\ may be artificial are not justified when the wide body of current knowledge about solar system minor bodies and planetary formation is considered. 

The Large Synoptic Survey Telescope (LSST) is expected to begin full operations in 2022 and is predicted to discover on the order of one interstellar object per year\cite{Cook:2016,Trilling:2017,Seligman:2018}. Thus, we will soon have a much better understanding of how common --- or rare --- the properties of \Ou\ are.  This knowledge will yield great insight into the planetesimal formation, evolution, and ejection processes at work across the Galaxy.
\\

\noindent{\bf THE \OU\ ISSI TEAM:}
\begin{hangparas}{.25in}{1}

\vspace{-0.1in}
Michele T. Bannister (Astrophysics Research Centre, Queen's University Belfast, Belfast BT7 1NN, UK) [ORCID: 0000-0003-3257-4490]

\vspace{-0.1in}
Asmita Bhandare (Max-Planck-Institut f\"ur Astronomie, K\"onigstuhl 17, 69117 Heidelberg, Germany)[ORCID: 0000-0002-1197-3946]

\vspace{-0.1in}
Piotr A. Dybczy\'nski (Astronomical Observatory Institute, Faculty of Physics, A. Mickiewicz University, S{\l}oneczna 36, Pozna\'n, Poland) [ORCID: 0000-0003-1492-4602]

\vspace{-0.1in}
Alan Fitzsimmons (Astrophysics Research Centre, Queen's University Belfast, Belfast BT7 1NN, UK) [ORCID: 0000-0003-0250-9911]

\vspace{-0.1in}
Aur{\'e}lie Guilbert-Lepoutre (Institut UTINAM, UMR 6213 / CNRS and Universit{\'e} de Bourgogne-Franche Comt{\'e}, F-25000 Besancon, France; Laboratoire de G{\'e}ologie de Lyon, LGL-TPE, UMR 5276 CNRS / Universit{\'e} de Lyon / Universit{\'e} Claude Bernard Lyon 1 / ENS Lyon, 69622 Villeurbanne, France) [ORCID: 0000-0003-2354-0766]

\vspace{-0.1in}
Robert Jedicke (Institute for Astronomy, University of Hawaii, 2680 Woodlawn Drive, Honolulu, HI 96822, USA) [ORCID: 0000-0001-7830-028X]

\vspace{-0.1in}
Matthew M. Knight* (University of Maryland, Department of Astronomy, College Park, MD 20742, USA) [ORCID: 0000-0003-2781-6897] 

\vspace{-0.1in}
Karen J. Meech (Institute for Astronomy, University of Hawaii, 2680 Woodlawn Drive, Honolulu, HI 96822, USA) [ORCID: 0000-0002-2058-5670]

\vspace{-0.1in}
Andrew McNeill (Department of Physics and Astronomy, Northern Arizona University, Flagstaff, AZ 86011, USA)

\vspace{-0.1in}
Susanne Pfalzner (Max-Planck-Institut f\"ur Radioastronomie, Auf dem H\"ugel 69, 53121 Bonn, Germany; J\"ulich Supercomputing Center, Forschungszentrum J\"ulich, 52428 J\"ulich, Germany; Department of Physics, University of Cologne, Z\"picher Str. 77, 50937 Cologne, Germany) [ORCID: 0000-0002-5003-4714]

\vspace{-0.1in}
Sean N. Raymond (Laboratoire d'Astrophysique de Bordeaux, CNRS and Universit{\'e} de Bordeaux, All{\'e}e Geoffroy St. Hilaire, F-33165 Pessac, France) [ORCID: 0000-0001-8974-0758]

\vspace{-0.1in}
Colin Snodgrass (Institute for Astronomy, University of Edinburgh, Royal Observatory, Edinburgh EH9 3HJ, UK) [ORCID: 0000-0001-9328-2905]

\vspace{-0.1in}
David E. Trilling (Department of Physics and Astronomy, Northern Arizona University, Flagstaff, AZ 86011 USA) [ORCID: 0000-0003-4580-3790]

\vspace{-0.1in}
Quanzhi Ye (Division of Physics, Mathematics and Astronomy, California Institute of Technology, Pasadena, CA 91125, USA; 
Infrared Processing and Analysis Center, California Institute of Technology, Pasadena, CA 91125, USA) [ORCID: 0000-0002-4838-7676]

*Contacting author: mknight2@umd.edu
\end{hangparas}

\noindent{\bf DATA AVAILABILITY}\\ 
The authors declare that the main data supporting the
findings of this study are available within the article and its Supplementary
Information files. Extra data are available from the corresponding author
upon request.\\

\noindent{\bf ACKNOWLEDGEMENTS}\\
We thank the International Space Science Institute (ISSI Bern), which made this collaboration possible. AF, MB and CS acknowledge support from UK Science and Technology Facilities Council grants ST/P0003094/1 and ST/L004569/1. KJM acknowledges support through NSF awards AST1617015, in addition to support for HST programs GO/DD-15405 and -15447 provided by NASA through a grant from the Space Telescope Science Institute, which is operated by the Association of Universities for Research in Astronomy under NASA contract NAS 5-26555. QY is supported by the GROWTH project funded by the National Science Foundation under Grant No.\ 1545949. This research was partially supported by the project 2015/17/B/ST9/01790 funded by the National Science Centre in Poland. MMK acknowledges support from NASA Near Earth Object Observations grant \#NNX17AK15G. AGL acknowledges funding from the European Research Council (ERC) under grant agreement No 802699. AM and DET are supported in part by Spitzer/NASA through an award issued by JPL/Caltech. SNR acknowledges helpful discussions with Phil Armitage related to the interstellar object number/mass density, and the Virtual Planetary Laboratory research team, funded by the NASA Astrobiology Program under NASA Grant Number 80NSSC18K0829. This work benefited from participation in the NASA Nexus for Exoplanet Systems Science research coordination network. 
\\

\noindent{\bf AUTHOR CONTRIBUTIONS}\\
MMK and AF organized the ISSI team. KJM created Figures~\ref{fig:res} \& \ref{fig:form}, SNR conducted the modelling of inferred interstellar object number density and created Figure~\ref{fig:num_density}. MMK, AF, and RJ created Figure~\ref{fig:elements} from source data provided by T.\ Engelhardt. All authors discussed the topics in the paper, contributed to the writing, and commented on the manuscript at all stages. 
\\

\noindent{\bf COMPETING INTERESTS}\\
The authors declare no competing financial interests.
\\

\noindent{\bf REFERENCES}
\\
\\
\bibliographystyle{naturemag}

\end{document}